\documentclass[final,journal]{IEEEtran}
\usepackage{amssymb,graphicx,color,cite}

\begin{document}

\title{A Pragmatic Coded Modulation Scheme for
High-Spectral-Efficiency Fiber-Optic Communications
\thanks{\noindent Benjamin P.\ Smith and 
Frank R.\ Kschischang are with the Electrical and Computer Engineering
Department, University of Toronto, 10 King's College Road, Toronto,
Ontario M5S 3G4, Canada  (e-mail:
\{ben, frank\}@comm.utoronto.ca).
}}

\author{Benjamin P. Smith, Frank R. Kschischang \IEEEmembership{Fellow, IEEE}} 

\IEEEpubid{0000--0000/00\$00.00~\copyright~2012 IEEE}

\maketitle

\begin{abstract}
\ifCLASSOPTIONonecolumn\relax\else\boldmath\fi
A pragmatic coded modulation system is presented that incorporates signal
shaping and exploits the excellent performance and efficient high-speed
decoding architecture of staircase codes.  Reliable communication within
$0.62$ bits/s/Hz of the estimated capacity (per polarization) of a system
with $L=2000$~km is provided by the proposed system, with an error floor
below $10^{-20}$. Also, it is shown that digital backpropagation increases
the achievable spectral efficiencies---relative to linear equalization---by
$0.55$ to $0.75$ bits/s/Hz per polarization.
\end{abstract}


\section{Introduction}

\IEEEPARstart{R}{ecent} progress has been made in estimating the
information-theoretic capacity of the class of fiber-optic communication
systems that are (presently) of commercial interest~\cite{EKWF2010a}, but
existing systems perform far from the fundamental limits of the channel.
While signal processing and coded modulation techniques promise to
eliminate this gap, their implementations---at the speeds present in
fiber-optic systems---present significant challenges.   

Many existing proposals for coded modulation in fiber-optic communication
systems amount to using techniques currently used in electrical wireline
and wireless communication systems. For example, in~\cite{mag2010}, the
authors propose a concatenated coding system with inner trellis-coded
modulation (for an 8-PSK constellation) and an outer product-like code, and
in~\cite{djor2010,djor2007,smi2010a}, the authors propose using low-density
parity-check (LDPC) codes for coded modulation. In both cases, the
proposals are verified by simulation, where the channel is assumed to be a
classical additive white Gaussian noise (AWGN), but no consideration is
given to the real-world implementation challenges for the proposed systems.  

Other proposals for coded modulation in fiber-optic systems consider
simplified channel models, and design codes for the resulting systems.  For
example, in~\cite{bop1995}, the authors design a trellis-coded
polarization-shift-keying modulation system, but their channel model only
considers laser phase noise, i.e., effects related to the propagation over
fiber are completely ignored.  In~\cite{bajk2010}, the authors consider a
nonlinear phase noise channel model studied by~\cite{lau07}, and design a
multi-level coded modulation system with Reed-Solomon codes at each level.
However, this channel model assumes a single-channel dispersion-less
system, which is not of practical interest. 

In this paper, we take a pragmatic approach to coded modulation for
fiber-optic systems, that addresses the deficiencies of the aforementioned
proposals. Due to the fact that product-like codes with syndrome-based
decoding have efficient high-speed decoders~\cite{Smi2011}, we consider
systems with hard-decision decoding. Furthermore, the channel model for
which the codes are designed is not a simplified one, but rather is derived
from (computationally intensive) simulations of the fiber-optic systems
based on the generalized nonlinear Schr\"odinger (GNLS) equation
[(\ref{eqn:gnls}), below], and thus accurately models the non-AWGN channel
that occurs in optical communication systems. 

In contrast to most classically studied communication channels, optical
fiber exhibits significant nonlinearity (in the intensity of the guided
light)~\cite{Agr2005a}. Furthermore, amplification acts as a source of
distributed AWGN, and fiber chromatic dispersion acts as a distributed
linear filter. Complicating matters, these three fundamental effects
\emph{interact} over the length of transmission.  In~\cite{EKWF2010a},
signal processing is performed via digital backpropagation, in order to
attempt to compensate the channel impairments.  However, their results do
not quantify the benefits of this compensation strategy.  Since digital
backpropagation is computationally expensive, one approach to reducing the
computational burden is to increase the step-size of the algorithm, as
in~\cite{ipk2008, ipk2010}.  In this paper, we compare the achievable rates
for two extreme cases: digital backpropagation (as in~\cite{EKWF2010a}),
and a linear equalizer (which can be considered as a form of ``linear
backpropagation'' in which the step-size is the system's length).

In Section~\ref{sec:pre}, we review staircase codes, the system model for a
fiber-optic communication system, and digital backpropagation.  In
Section~\ref{sec:rates}, we compare the transmission rates that can be
achieved using digital backpropagation with those achievable by linear
equalization. In Section~\ref{sec:cmod}, we present the details of a
pragmatic coded modulation system, and compare the performance of the
system to the capacity estimates.

\IEEEpubidadjcol

\section{Preliminaries} \label{sec:pre}

\subsection{Staircase Codes} \label{sec:sc}

Staircase codes~\cite{Smi2011} are a family of high-rate binary
error-correcting codes suitable for high-speed fiber-optic communications.
Staircase codes can be interpreted as \emph{generalized} LDPC codes, that
is, sparse graph-based codes whose constraint nodes are
error-\emph{correcting} codes, not the single-parity-check
error-\emph{detecting} codes used in the constraint nodes of conventional
LDPC codes.  With such generalized LDPC codes, algebraic decoding can be
applied at the constraint nodes, and the decoder can operate exclusively on
syndromes. As discussed in \cite{Smi2011}, this significantly reduces the
decoder data-flow (relative to a message-passing LDPC decoder), admitting
an efficient high-speed implementation.  Furthermore, due to the error
correcting capabilities of the constraint nodes, staircase codes have very
low error floors, which can be estimated analytically.  Finally, due to
their structural properties, staircase codes provide superior performance
to product codes.

Staircase codes are completely characterized by the relationship between
successive matrices of symbols. Specifically, consider the (infinite)
sequence $B_0,B_1,B_2,\ldots$ of $m$-by-$m$ matrices $B_i$, $i \in
\mathbb{Z}^{+}$. Block $B_0$ is initialized to a reference state known to
the encoder-decoder pair, e.g., block $B_0$ could be initialized to the
all-zeros state, i.e., an $m$-by-$m$ array of zero symbols.  Furthermore,
we select a conventional FEC code (e.g., Hamming, BCH, Reed-Solomon, etc.)
in systematic form to serve as the \emph{component} code; this code, which
we henceforth refer to as $C$, is selected to have blocklength $2m$
symbols, $r$ of which are parity symbols.   

Generally, the relationship between successive blocks in a staircase code
satisfies the following relation: for any $i\geq 1$, each of the rows of
the matrix $\left[ B_{i-1}^T B_i \right]$ is a valid codeword in $C$.  Just
as in a conventional product code, any given symbol in any given block
$B_i$ participates in two constraints:  one to satisfy the condition that
each row of $\left[  B_{i-1}^T B_i \right]$ is a codeword of $C$, and one
to satisfy the condition that each row of $\left[ B_{i}^T B_{i+1} \right]$
is a codeword of $C$.

\subsection{System Model} \label{sec:sys}

We consider a coherent fiber-optic communication system.  Between the
transmitter and receiver, standard-single-mode fiber and ideal distributed
Raman amplification are assumed, but we note that the methods presented
herein also apply to alternate system configurations (e.g., systems with
inline dispersion-compensating fiber, and/or lumped amplification). The
complex baseband representation of the signal in a single polarization at
the output of the transmitter is $A(0,t)$, and at the input of the receiver
is $A(L,t)$, where $L$ is the total system length; note that $A(z,t)$
represents the full field, i.e., in general it represents co-propagating
dense wavelength-division-multiplexed signals. 

The generalized non-linear Schr\"{o}dinger (GNLS) equation expresses the
evolution of $A(z,t)$:
\begin{equation}
\frac{\partial A}{\partial z} +
\frac{j\beta_2}{2}\frac{\partial^2 A}{\partial t^2} -
j\gamma|A|^2 A=n(z,t).
\label{eqn:gnls}
\end{equation}
Since ideal distributed Raman amplification is assumed, the loss term has
been omitted, and $n(z,t)$ is a circularly symmetric complex Gaussian noise
process with autocorrelation
\[
\mathcal{E}\left[n(z,t) n^{\star}(z',t')\right] =
\alpha h v_s K_T \delta(z-z',t-t'),
\]   
where $h$ is Planck's constant, $v_s$ is the optical frequency, and $K_T$
is the phonon occupancy factor.  In Table~\ref{tab:param}, we provide
parameter values for the system components.

Note that the scalar equation (\ref{eqn:gnls})---whose numerical solution
is used to generate all of the results of this paper---governs propagation
of waveforms in a single polarization mode.  The achievable rates for a
dual-polarized transmission system would be approximately (but slightly
less than) twice as large as for the single polarization system considered
here, but a more complicated vector version of (\ref{eqn:gnls}), taking
into account the effects of fiber birefringence and coupling between the
polarization modes as well as the stochastic nature of polarization mode
dispersion, would need to be considered.

\begin{table}[t]
\centering
\caption{System Parameter Values}
\begin{tabular}{ll} \hline
Second-order dispersion $\beta_2$ & -21.668 $\mathrm{ps}^2/\mathrm{km}$ \\
Loss $\alpha$ & $4.605\times 10^{-5}$ $\mathrm{m}^{-1}$  \\ 
Nonlinear coefficient $\gamma$ & $1.27$ $\mathrm{W}^{-1}\mathrm{km}^{-1}$ \\
Center carrier frequency $v_s$ &  $193.41$ THz \\ 
Phonon occupancy factor $K_T$ & 1.13\\ \hline
\end{tabular}
\label{tab:param}
\end{table}

\subsection{Digital Backpropagation}

Throughout propagation over an optical fiber, stochastic effects (noise),
linear effects (dispersion) and nonlinear effects (Kerr nonlinearity)
\emph{interact}, and---even in the absence of noise---solving the GNLS
equation requires numerical techniques. On the other hand, in the absence
of noise, the system is invertible, i.e., the transmitted signal $A(0,t)$
can be recovered from the received signal $A(N L_A)$ by inverting the
channel.  When the channel is inverted by digital signal processing, we say
the receiver performs \emph{digital backpropagation}. 

The most commonly used numerical method to solve the GNLS equation is the
split-step Fourier method~\cite{A2006a,SHZ2003a}. The basic idea is to
divide the total fiber length into short segments, then to consider each
segment as the concatenation of (separable) nonlinear  and linear
transforms (for distributed amplification, an additive noise is added after
the linear step). In the following, we briefly review the split-step
Fourier method. For simplicity of the presentation, we ignore the effects
of amplification, which can be incorporated into a numerical solver in an
obvious manner.

For a known $A(z=z_0,t)$, the split-step Fourier method calculates
$A(z=z_0+h,t)$ as follows. First, in the absence of linear effects, the
GNLS equation has the form,
\[
\frac{\partial A}{\partial z}=j\gamma|A|^2 A,
\]
with solution,
\[
A(z=z_0+h,t)=A(z=z_0,t)\exp(j \gamma |A(z=z_0,t)|^2 h).
\]
We now use this solution as the input to the the linear step, i.e., let 
\[
\hat{A}(z=z_0,t)=A(z=z_0,t)\exp(j \gamma |A(z=z_0,t)|^2 h)
\]
be the input to the linear step. The linear form of the GNLS equation is
\[
\frac{\partial A}{\partial z}=-\frac{\alpha}{2}A-\frac{j\beta_2}{2}\frac{\partial^2 A}{\partial t^2},
\]
which can be efficiently solved in the frequency domain. Defining
\[
A(z,t)=\frac{1}{2\pi} \int_{-\infty}^{\infty} \tilde{A}(z,\omega) \exp(j\omega t) d\omega,
\]
it can be shown that 
\begin{equation}
\tilde{A}(z=z_0+h,\omega)=\tilde{A}(z=z_0,\omega)\exp\left(\left(j\frac{\beta_2}{2}\omega^2-\frac{\alpha}{2}\right)h\right).
\label{eq:linstep}
\end{equation}
Putting this together, we have
\begin{IEEEeqnarray*}{rCl}
A(z=z_0+h,t)&=&\mathcal{F}^{-1}\left\{\mathcal{F}\left\{\hat{A}(z=z_0,t)\right\} \right.  \\
& & \mbox{} \cdot \left. \exp\left(\left(j\frac{\beta_2}{2}\omega^2-\frac{\alpha}{2}\right)h\right)\right\},
\end{IEEEeqnarray*}
where $\mathcal{F}$ is the Fourier transform operator.

Digital backpropagation is then accomplished by the split-step Fourier
method, using a negative step-size $h$.  Note that, in general, $A(z,t)$ is
the complex envelope of a multi-channel optical signal.  It follows that
\emph{full compensation} of channel impairments---even if only a single
channel is of interest to the receiver---requires backpropagation to be
performed on the \emph{multi}-channel signal, since nonlinearity induces
interaction between signal components at non-overlapping frequencies.
However, in practice, receivers operate on a per-channel basis. Even if a
multi-channel receiver were available, co-propagating channels may be
optically-routed in or out throughout transmission, and thus channels that
have co-propagated with the desired channel may not even be available at
the receiver (and those channels that are available may not have
co-propagated with the desired channel). Therefore, we consider
single-channel backpropagation, in which the receiver first extracts the
channel of interest from $A(z=L,t)$ (via a bandpass filter), and performs
digital backpropagation on the corresponding signal. 

\section{Achievable Rates} \label{sec:rates}

Although many current state-of-the-art systems include some form of
electronic dispersion compensation (i.e., equalization) in the receiver,
digital backpropagation is significantly more computationally intensive,
since many steps---each of which has roughly the complexity of a standard
equalization scheme---of the split-step Fourier method are required to
accurately compensate the nonlinear effects.

In this section we compare the achievable information rates when (only)
linear equalization is performed to the achievable rates of a system that
performs digital backpropagation.  Furthermore, the resulting capacity
estimates serve as upper bounds on the performance of a coded modulation
system, the design of which we consider in Section~\ref{sec:cmod}.

\subsection{Memoryless Capacity Estimation}

In~\cite{EKWF2010a}, Essiambre et al.\ present an estimate of the
information theoretic capacity of optical fiber networks. In this section,
we review their technique, which we will make use of in the following.  

\subsubsection{Transmitter}
We consider a system that employs pulse-amplitude modulation (PAM) with
(orthonormal) sinc pulses. That is, the transmitted signal (corresponding
to the baseband representation of the $l$-th channel) is of the form
\[
X_l(t)=\sum_{k=-\infty}^{\infty}  \frac{\phi_{k,l}}{\sqrt{T_s}} {\rm sinc}\left(\frac{t-kT_s}{T_s}\right),
\]
where ${\rm sinc}(\theta)=\frac{\sin \pi \theta}{\pi \theta}$.  The
$\phi_{k,l}$ are elements of a discrete-amplitude continuous-phase input
constellation $\mathcal{M}$, i.e, for $N$ rings, $\theta \in [0,2\pi)$, and
$r\geq0$, 
\[
\mathcal{M}=\left\{m \cdot r\exp \left(j\theta\right)|m \in \{1,2,\ldots,N\}\right\}.
\]
Each ring is assumed equiprobable, and for a given ring, the phase
distribution is uniform.  This choice of constellation is motivated by the
fact that the channel represented by the GNLS equation can be argued to be
statistically rotationally invariant (i.e., for a channel with conditional
distribution $f(y|x)$, $f(y|x_0)=f(y \exp \left( j\theta \right)|x_0 \exp
\left(j\theta \right))$ for $\theta \in [0,2\pi)$) and thus points on the
same ring can be considered ``equivalent'', which reduces the computational
requirements in characterizing the channel.  Furthermore, it is well known
that, for sufficiently many rings, the Shannon Limit of the AWGN channel
can be closely approached, and one would expect this to be true also for
the non-AWGN channel considered here.

In the general case of a multi-channel system having $2B+1$ channels with a
channel spacing $1/T_s$ Hz, the input to the fiber has the form
\[
A(z=0,t)=\sum_{k=-\infty}^{\infty} \sum_{l=-B}^B  \frac{\phi_{k,l}}{\sqrt{T_s}} {\rm sinc}\left(\frac{t-kT_s}{T_s}\right) e^{j 2\pi l t/T_s }.
\] 

\subsubsection{Receiver}
\begin{figure}[t]
\centering
\ifCLASSOPTIONonecolumn
\includegraphics[width=0.5\columnwidth]{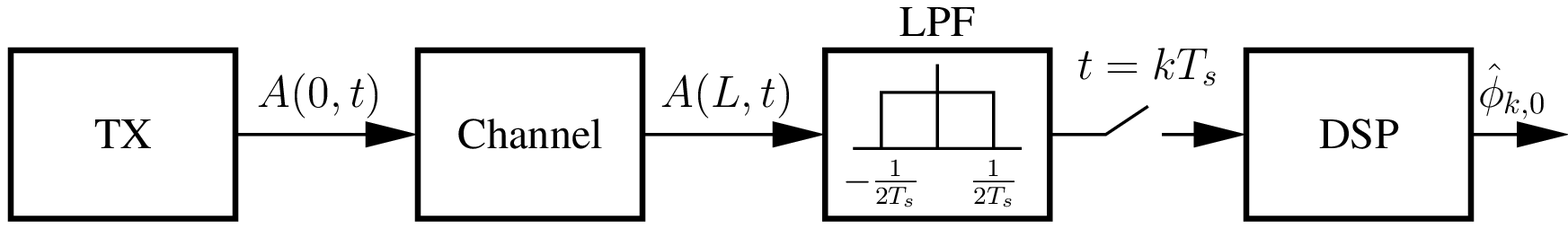}
\else
\includegraphics[width=\columnwidth]{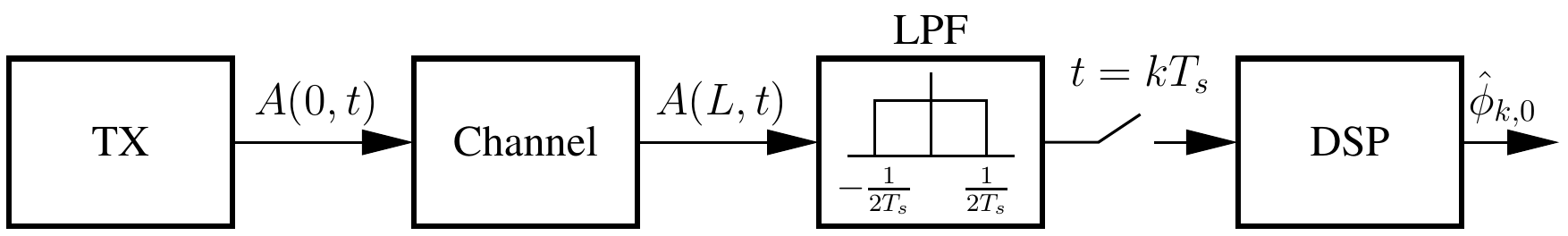}
\fi
\caption{System model for memoryless capacity evaluation} \label{fig:sysmem}
\end{figure}

By convention, the channel of interest (COI) is assumed to correspond to
$l=0$.  From the channel output $A(L,t)$, the (baseband) digital coherent
optical receiver extracts the COI via an ideal low-pass filter, and the
corresponding signal is sampled at the rate $1/T_s$. The resulting
discrete-time signal is then compensated by digital signal processing,
i.e., backpropagation (BP) or linear equalization (EQ), providing estimates
$\hat{\phi}_{k,0}$ of the transmitted symbols $\phi_{k,0}$, as illustrated
in Fig~\ref{fig:sysmem}.

\subsubsection{Channel Model}
In order to facilitate the capacity estimation, the discrete-time channel
is assumed to be \emph{memoryless}, i.e., it is assumed that
backpropagation removes any dependence (introduced by the channel) between
received symbols.  The (memoryless) conditional distribution of the channel
is estimated from numerical simulations. 

\begin{figure}[t]
\centering
\ifCLASSOPTIONonecolumn
\includegraphics[width=0.5\columnwidth]{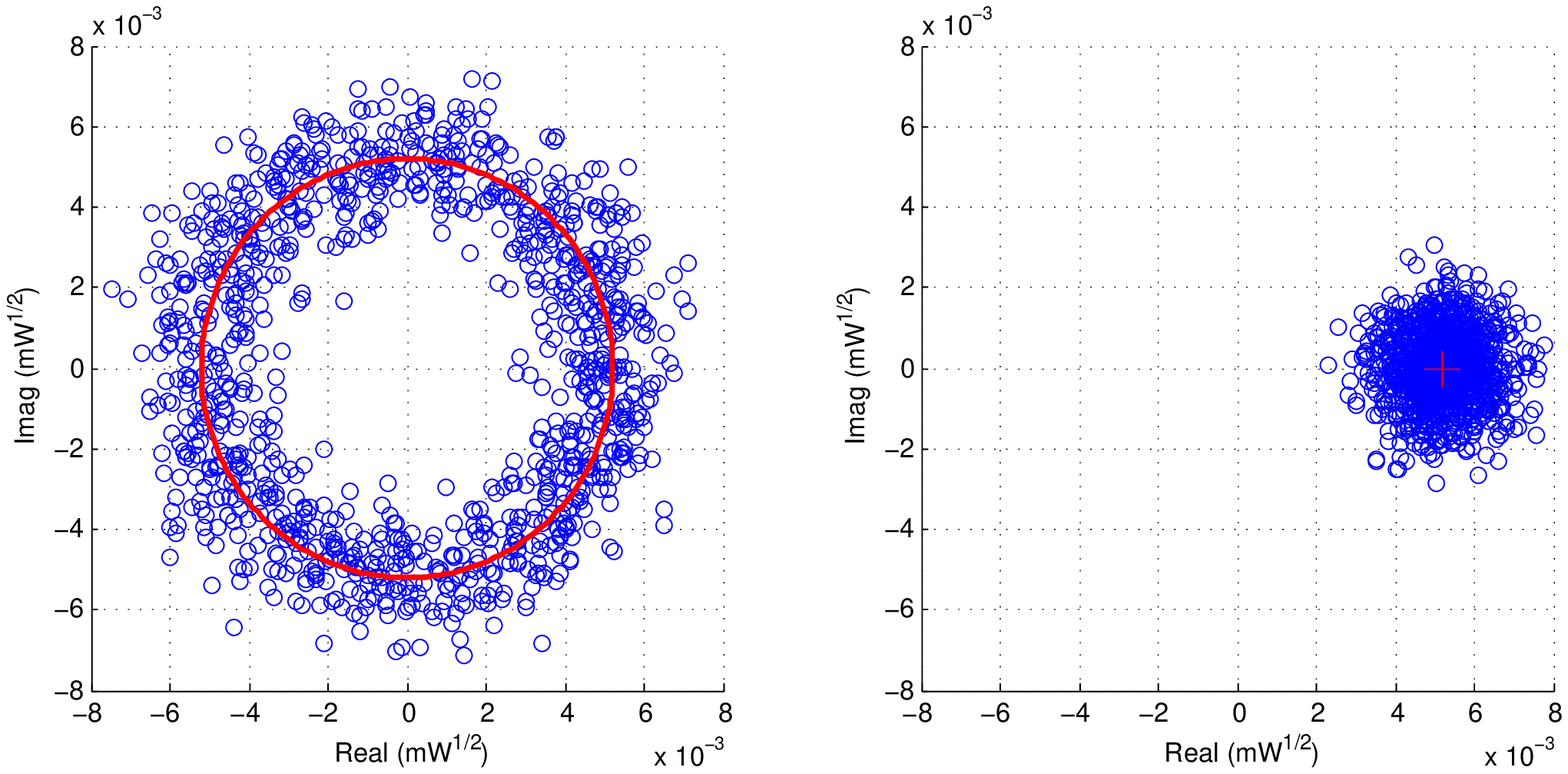}
\else
\includegraphics[width=\columnwidth]{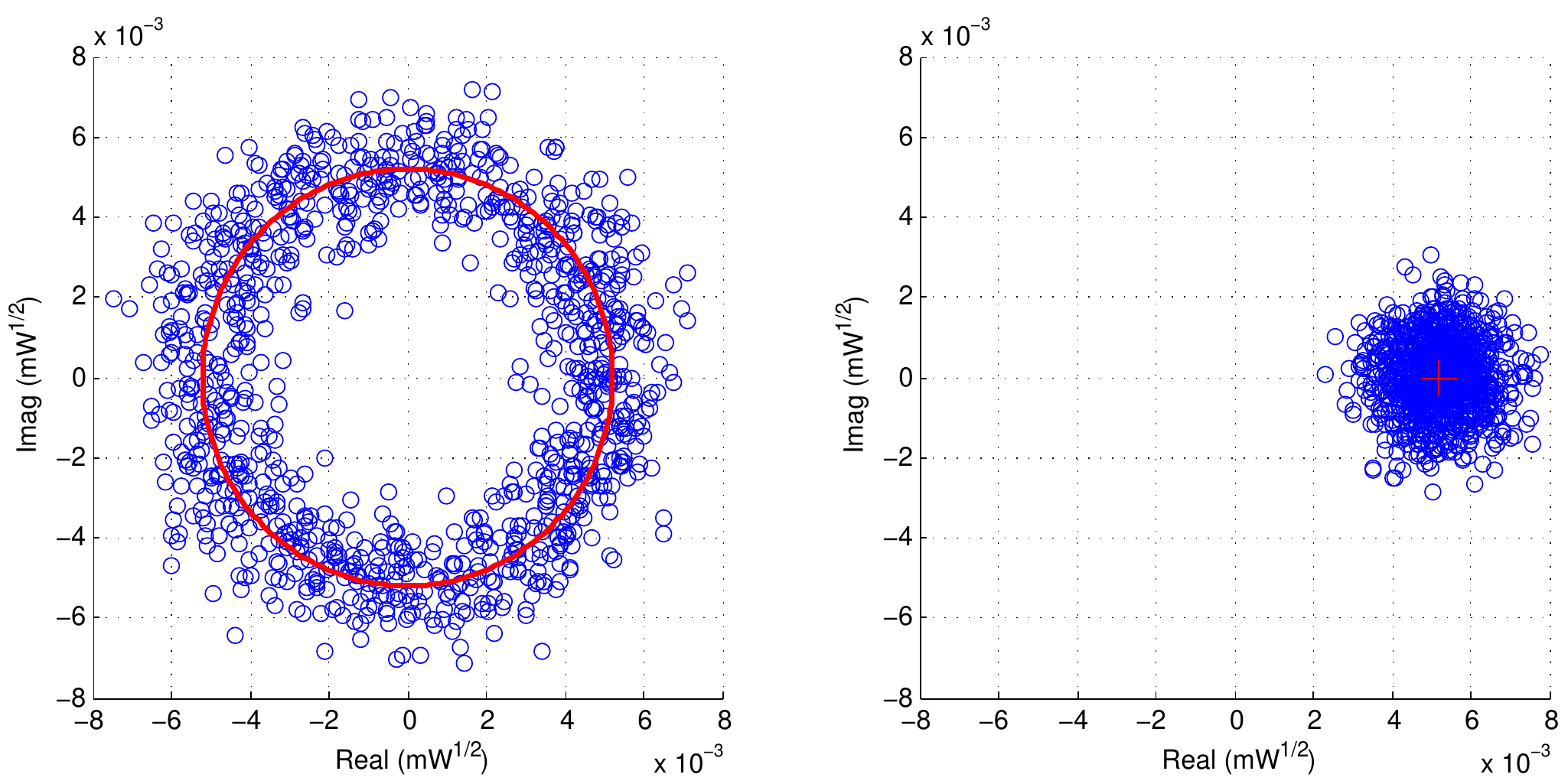}
\fi
\caption{Channel outputs for a fixed-ring input, and back-rotated outputs.} \label{fig:backrot}
\end{figure}

Since the channel is statistically rotationally invariant, observations of
transmitted points from the same ring are first `back-rotated' to the real
axis, as illustrated in Fig.~\ref{fig:backrot}.  The back-rotated points
are represented by $\tilde{\phi}_{k,l}$,
\[
\tilde{\phi}_{k,l}=\hat{\phi}_{k,l}
\exp\left(-j ( \Phi_{\rm XPM}+ \angle \phi_{k,l}) \right),
\]
where $\Phi_{\rm XPM}$ is a constant (input-independent) phase rotation
contributed by cross-phase modulation (XPM).

Next, for each $i$ and a fixed $l$ (the channel of interest), we calculate
the mean $\mu_i$ and covariance matrix $\Omega_i$  (of the real and
imaginary components) of those $\tilde{\phi}_{k,l}$ corresponding to the
$i$-th ring, and model the distribution of those $\tilde{\phi}_{k,l}$ by
$\mathcal{N}(\mu_i,\Omega_i)$.  Finally, from the rotational invariance of
the channel, the channel is modeled as
\[
f\left(y|x=r\cdot i  \exp \left( j \phi \right) \right) \sim \mathcal{N}(\mu_i \exp \left( j \phi \right),\Omega_i),
\]
where the (constant) phase rotation due to $\Phi_{\rm XPM}$ is ignored,
since it can be canceled in the receiver.  Note that this model reduces to
an additive `noise' model when $\mu_i=(r\cdot i,0)$, but in general this
relationship need not be true. 

\subsubsection{Capacity Estimation}
The mutual information of the memoryless channel is
\[
I(X;Y)=\int \! \! \int \! f(x,y) \log_2 \frac{f(y|x)}{f(y)}\,dx\,dy,
\]
where $f(x)$ represents the input distribution on $\mathcal{M}$ with
equiprobable rings and a uniform phase distribution, which provides an
estimate of the capacity of an optically-routed fiber-optic communication
system.

\subsubsection{Signaling Parameters}
In Table~\ref{tab:paramsign} we provide the parameters of the signaling
scheme, to be used throughout the remainder of this work. In general,
further increasing the number of simulated channels has a negligible effect
on the capacity estimates.
\begin{table}[ht]
\centering
\caption{Signaling parameter values}
\begin{tabular}{ll} \hline
Baud rate $1/T_{s}$ & 100 GHz \\ 
Channel bandwidth $W$ & 101 GHz \\ 
Number of rings $N$ & 64 \\ 
Number of channels  & $2B+1=5$ \\ \hline
\end{tabular}
\label{tab:paramsign}
\end{table}

\subsection{Results}
\begin{figure}[t]
\centering
\ifCLASSOPTIONonecolumn
\includegraphics[width=0.5\columnwidth]{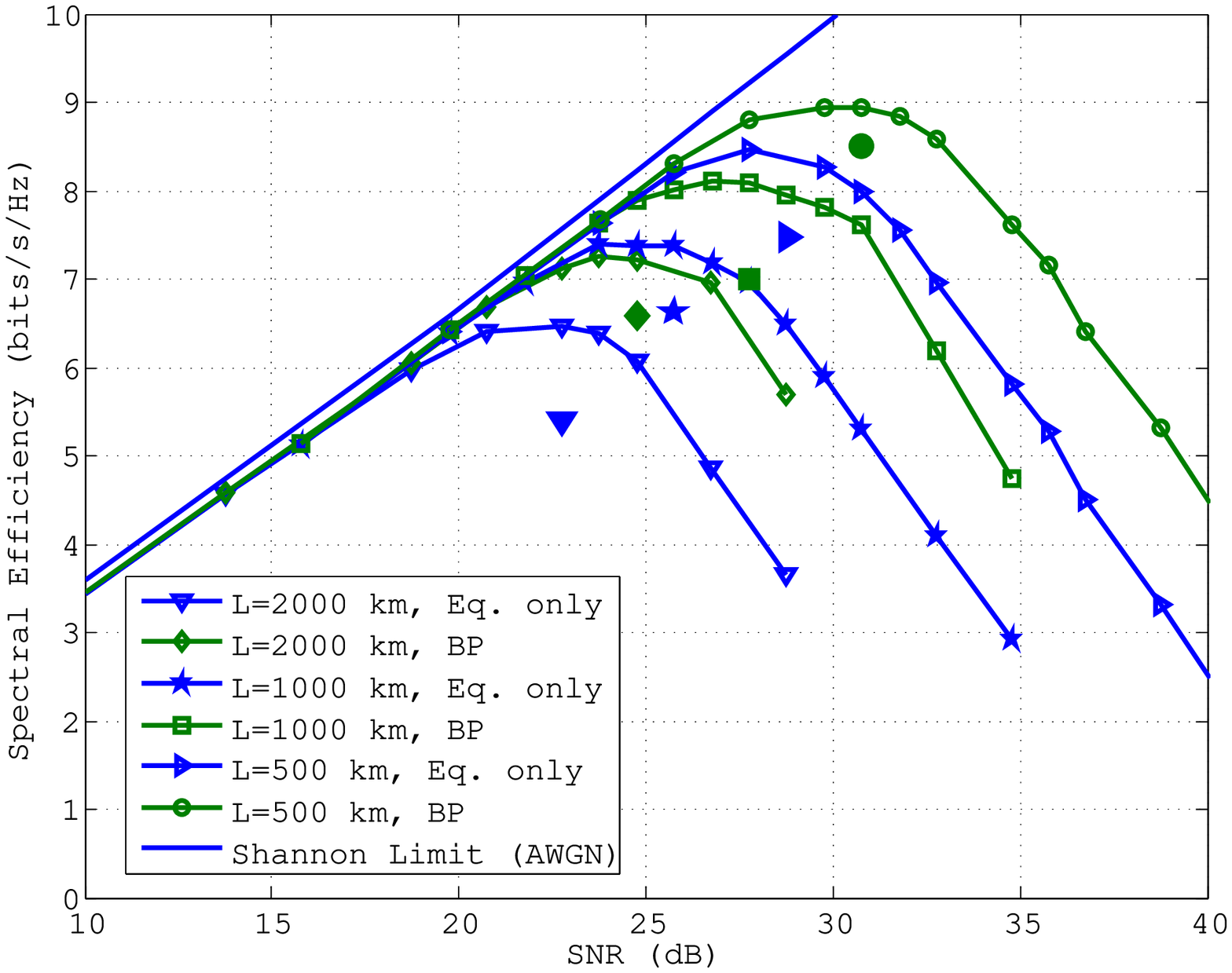}
\else
\includegraphics[width=0.950\columnwidth]{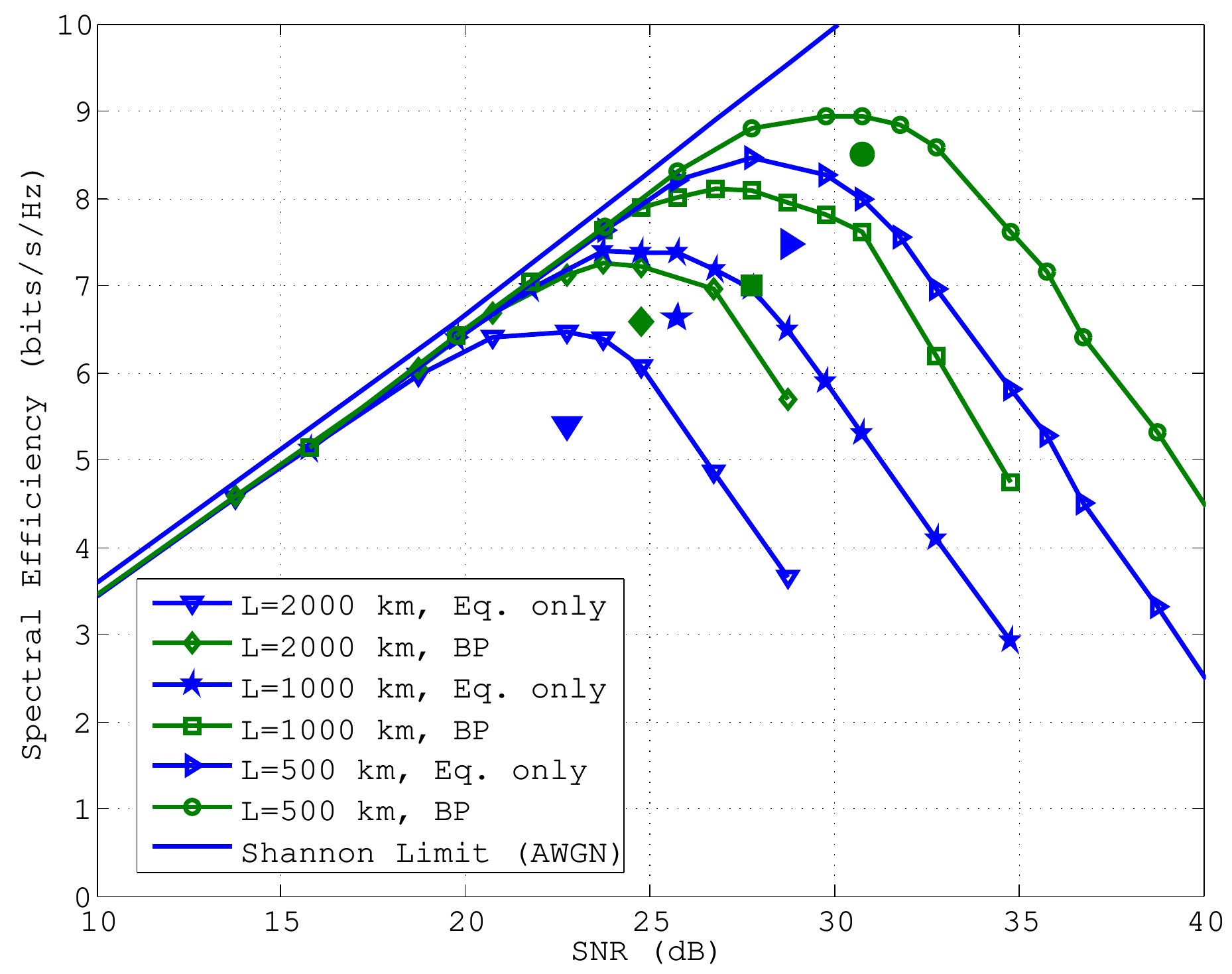}
\fi
\caption{(Theoretically) achievable spectral 
efficiencies for BP and EQ at different transmission lengths.  Also 
shown (by the isolated symbols) are the spectral efficiencies achieved by the staircase-coded
systems described in Table~\ref{tab:scbicm}.} \label{fig:ramwdm}
\end{figure}

In Fig.~\ref{fig:ramwdm}, we present the achievable spectral efficiencies.
We consider systems of length $L=500$, $1000$ and $2000$~km. The
signal-to-noise ratio (SNR) is defined as
\[
{\rm SNR}=\frac{P}{N_{\rm ASE} W},
\]
where $P$ is the average transmitter power, $W$ is the bandwidth occupied
by a single channel, and $N_{\rm ASE}=L \alpha h v_s K_T$ is the power
spectral density of the noise.  In contrast to conventional linear Gaussian
channel models, $N_{\rm ASE}$ is fixed by the choices of $L$ and the
amplification technique. Therefore, for a fixed system, the SNR can be
increased only by increasing the input power.
 
For $L=2000$~km, the peak spectral efficiency is approximately $6.45$
bits/s/Hz per polarization when only linear equalization is performed, but
increases to approximately $7.2$ bits/s/Hz per polarization for digital
backpropagation.  For $L=1000$~km, the peak spectral efficiency is
approximately $7.4$ bits/s/Hz per polarization when only linear
equalization is performed, but increases to approximately $8.1$ bits/s/Hz
per polarization for digital backpropagation. Finally, for $L=500$~km, the
peak spectral efficiency is approximately $8.45$ bits/s/Hz per polarization
when only linear equalization is performed, but increases to approximately
$9.0$ bits/s/Hz per polarization for digital backpropagation.

For the cases considered, digital backpropagation increases the achievable
spectral efficiencies---relative to linear equalization---by $0.55$ to
$0.75$ bits/s/Hz per polarization. From the standpoint of achievable rates,
the channel is ``nearly'' linear for most input powers of interest, in the
sense that linear equalization achieves rates that closely approach those
achievable via backpropagation.  Furthermore, even when the input power is
such that the achievable rate is maximized, the distortion introduced by
the channel is well-modeled as AWGN, and thus classical coding methods
ought to provide near-capacity reliable communications.  However, due to
the extremely high per-channel data rates of fiber-optic systems,
implementation challenges arise. In the following, we propose a pragmatic
coded modulation system---based on staircase codes---that provides
excellent performance and an efficient high-speed implementation.

\section{A Pragmatic Coded-Modulation Scheme} \label{sec:cmod}

Although staircase codes are binary error-correcting codes with a
syndrome-based decoding algorithm, they can be adapted---via known
techniques---to provide error-correction in high-spectral-efficiency
communication systems, while maintaining their efficient decoding
architecture.  In the following, we first review these techniques, then we
provide the parameters of staircase-coded systems and present their
performance. 

\subsection{Coding}

For high-spectral-efficiency communications, the set of channel input
symbols (i.e., the modulation constellation) must be sufficiently large,
and coding is required on the resulting non-binary input alphabet. At first
glance, this would seem to require the design of error-correction codes
over non-binary alphabets, with a decoding algorithm that accounts for the
distance metric implied by the underlying channel.  Indeed, this `direct'
approach provides motivation for trellis-coded modulation~\cite{U1982a}, in
which the code is designed to optimize the minimum Euclidean distance
between transmitted sequences. 

Alternatively, by considering the set of channels induced by the bit-labels
of the constellation points, coded modulation via binary codes can be
applied with---in principle---no loss of optimality. To see this, consider
a $2^M$-point constellation $\mathcal{A}$, for which each symbol is labeled
with a unique binary $M$-tuple $(b_1,b_2,\ldots,b_M)$.  For a channel with
input $X \in \mathcal{A}$, and an output $Y$, the capacity of the resulting
channel is $I(X;Y)$ (maximized over the input distribution $p(x)$), which
can be expanded by the chain rule of mutual information:
\begin{IEEEeqnarray}{rCl}
I(X;Y)&=&I(b_1,b_2,\ldots,b_M;Y) \nonumber \\
&=&I(b_1;Y)+I(b_2;Y|b_1)+\cdots+ \nonumber \\
& & I(b_M;Y|b_1,b_2,\ldots,b_{M-1}) \label{eq:mlc} 
\end{IEEEeqnarray}
Note that each term (i.e., the sub-channels) in the expansion defines a
binary-input channel, for which binary error-correction codes---such as
staircase codes---can be applied; this approach is referred to as
multi-level coding (MLC)~\cite{WFH1999a}. Furthermore, if a
capacity-approaching code is applied to each sub-channel, then the capacity
of the modulation scheme is achieved, that is, there is no loss in
optimality in applying binary coding to each sub-channel.  However, from
(\ref{eq:mlc}), it is implied that decoding is performed in stages, since
decoded bits from lower-indexed levels provide side information for
decoding higher levels; the resulting decoding architecture is referred to
as a multi-stage decoder. 

\begin{figure}[t]
\centering
\ifCLASSOPTIONonecolumn
\scalebox{0.66}{\includegraphics{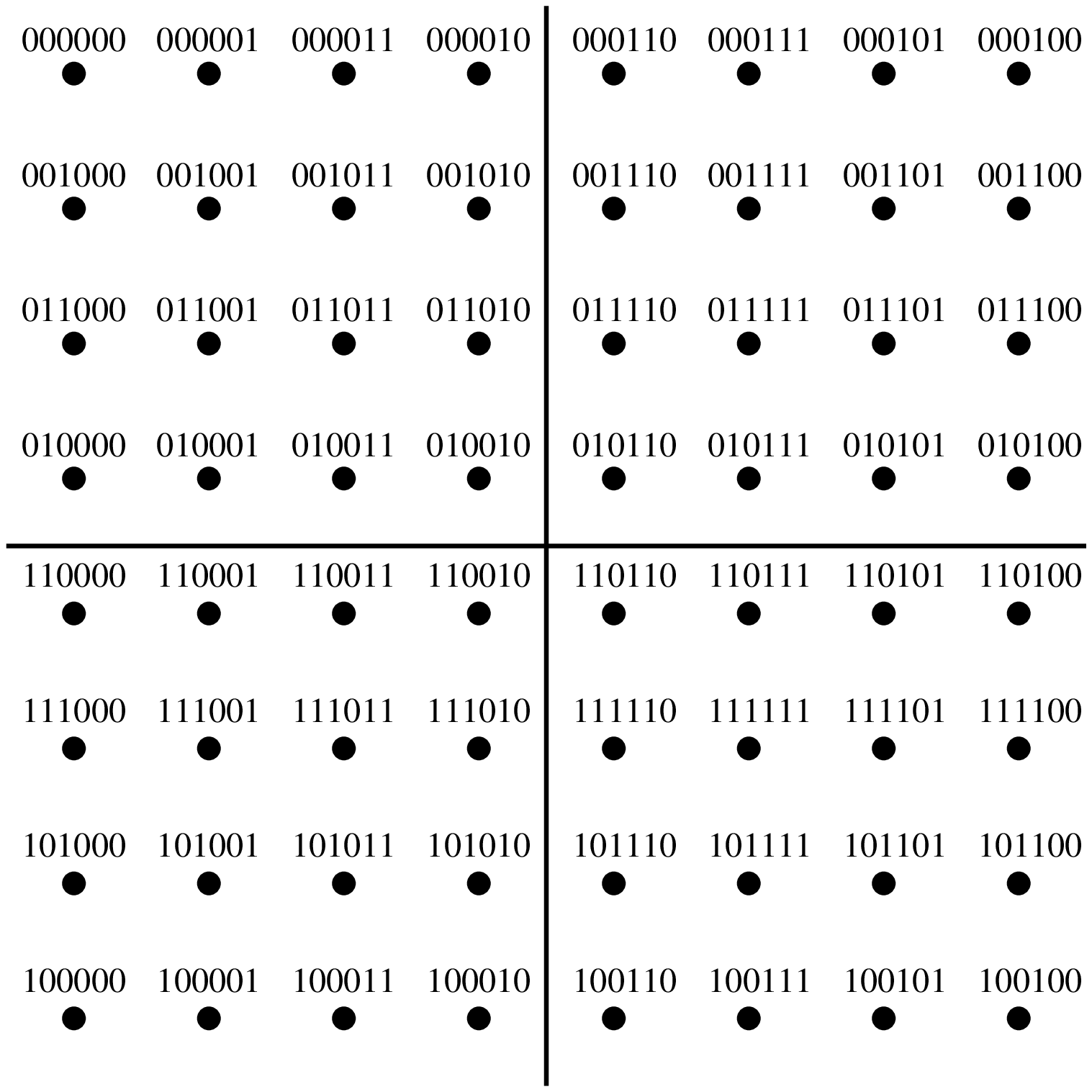}}
\else
\scalebox{0.33}{\includegraphics{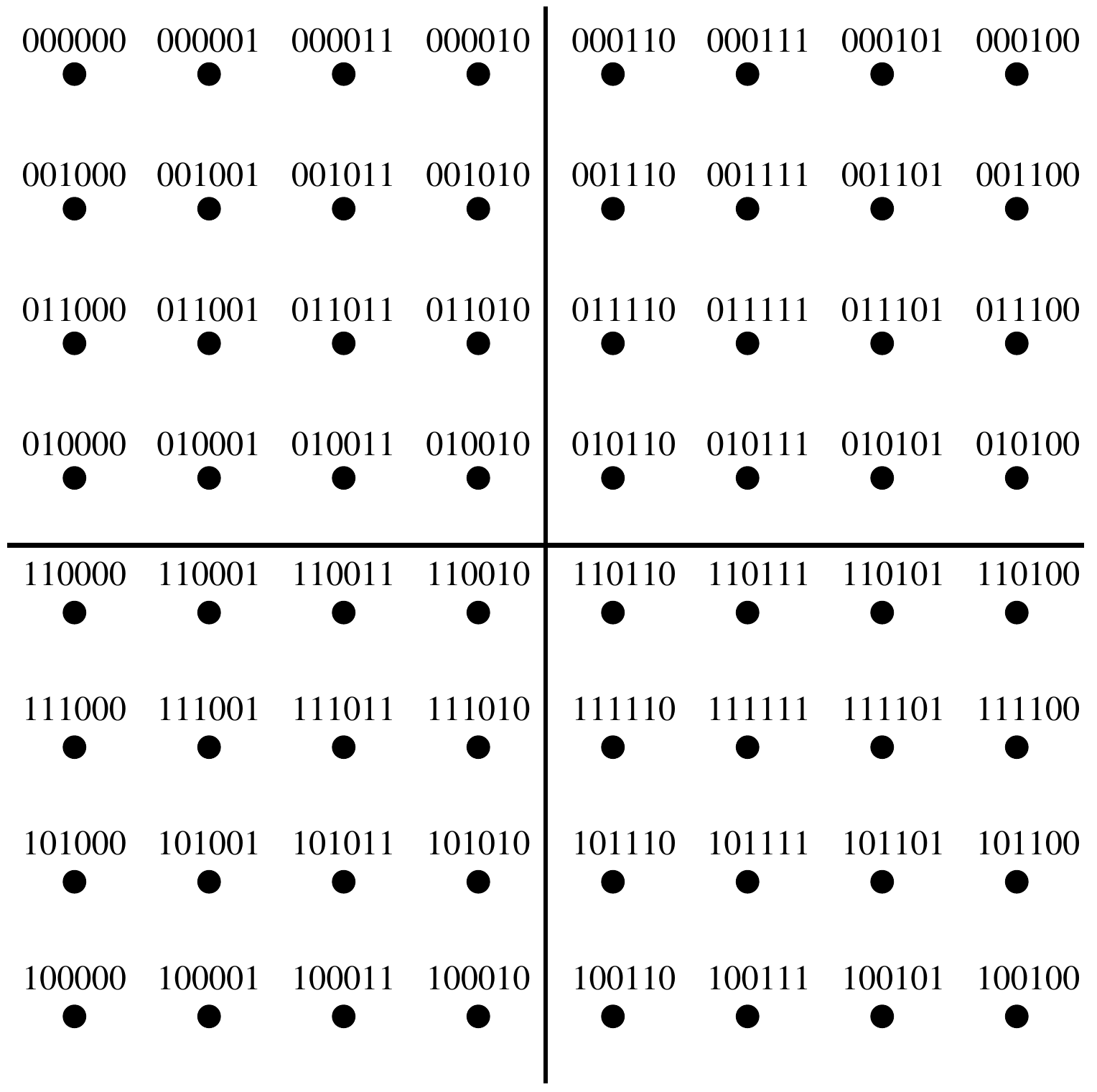}}
\fi
\caption{A Gray-labeled 64-QAM constellation.} \label{fig:64}
\end{figure}

Note that the multi-stage architecture introduces decoding latency to the
higher levels, requires memory to store channel outputs prior to decoding
(since outputs are `held' until decoded bits from the lower levels are
available), and requires an individual code for each sub-channel. Clearly,
the latency and memory issues can be eliminated simply by ignoring the
conditioning in (\ref{eq:mlc}), and the resulting system has capacity
\[
C_{\rm PID}=I(b_1;Y)+I(b_2;Y)+\cdots+I(b_M;Y),
\]
where PID stands for ``parallel independent decoding''. However, even when
capacity-achieving codes (i.e., with rates $I(b_i;Y)$) are applied to each
sub-channel, $C_{\rm PID}$ may be significantly less than $I(X;Y)$.  Note
that the capacities of the individual bit-channels depend on the
constellation labeling; for MLC their overall sum is fixed, regardless of
the labeling, but for PID their sum (i.e., $C_{\rm PID}$) depends on the
labeling. In fact, for Gray-labeling\footnote{A Gray-labeling has the
property that the binary $M$-tuples of nearest neighbor constellation
points differ in only a single position}  (see Fig.~\ref{fig:64}), the
difference between $C_{\rm PID}$ and  $I(X;Y)$ essentially vanishes, as
shown in~\cite{CTB1998a}. Furthermore, even though the capacities of the
individual sub-channels are not identical, a single binary error-correcting
code (whose rate is the average of the bit-channel rates) provides
near-capacity performance, which addresses the third issue with MLC; this
approach is referred to as bit-interleaved coded modulation (BICM).

\subsection{Shaping}

Implicit in the definition of channel capacity is an optimization over the
input alphabet of the channel. For example, the optimal input distribution
for the additive white Gaussian noise channel is itself Gaussian.  Indeed,
the ring-like constellations used in Section~\ref{sec:rates} provide
shaping gain relative to a QAM-like constellation.  In order to approach
the capacity estimates of the fiber optic channel, shaping is essential in
any coded modulation scheme. 

\begin{figure}[t]
\centering
\ifCLASSOPTIONonecolumn
\scalebox{0.66}{\includegraphics{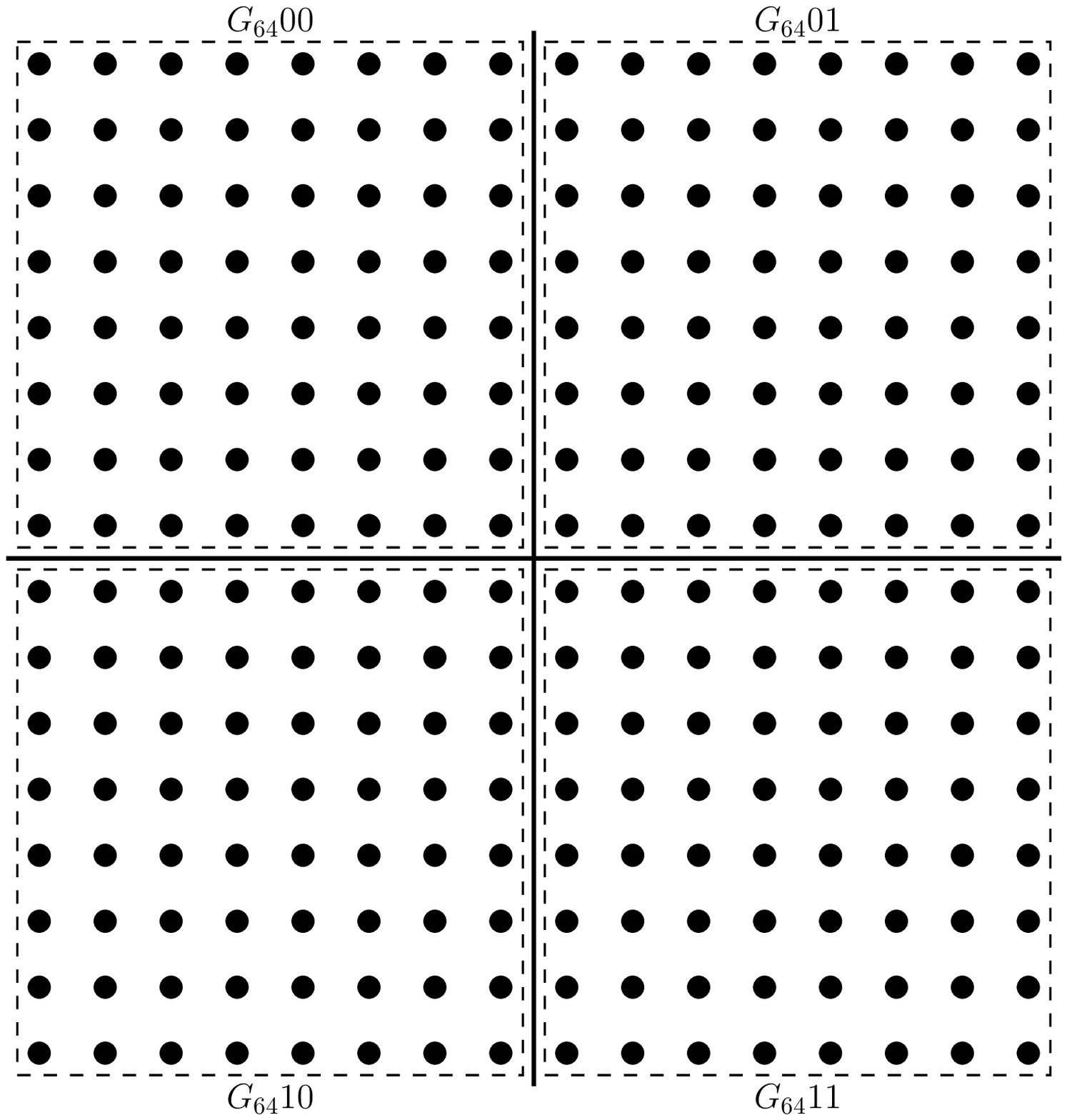}}
\else
\scalebox{0.33}{\includegraphics{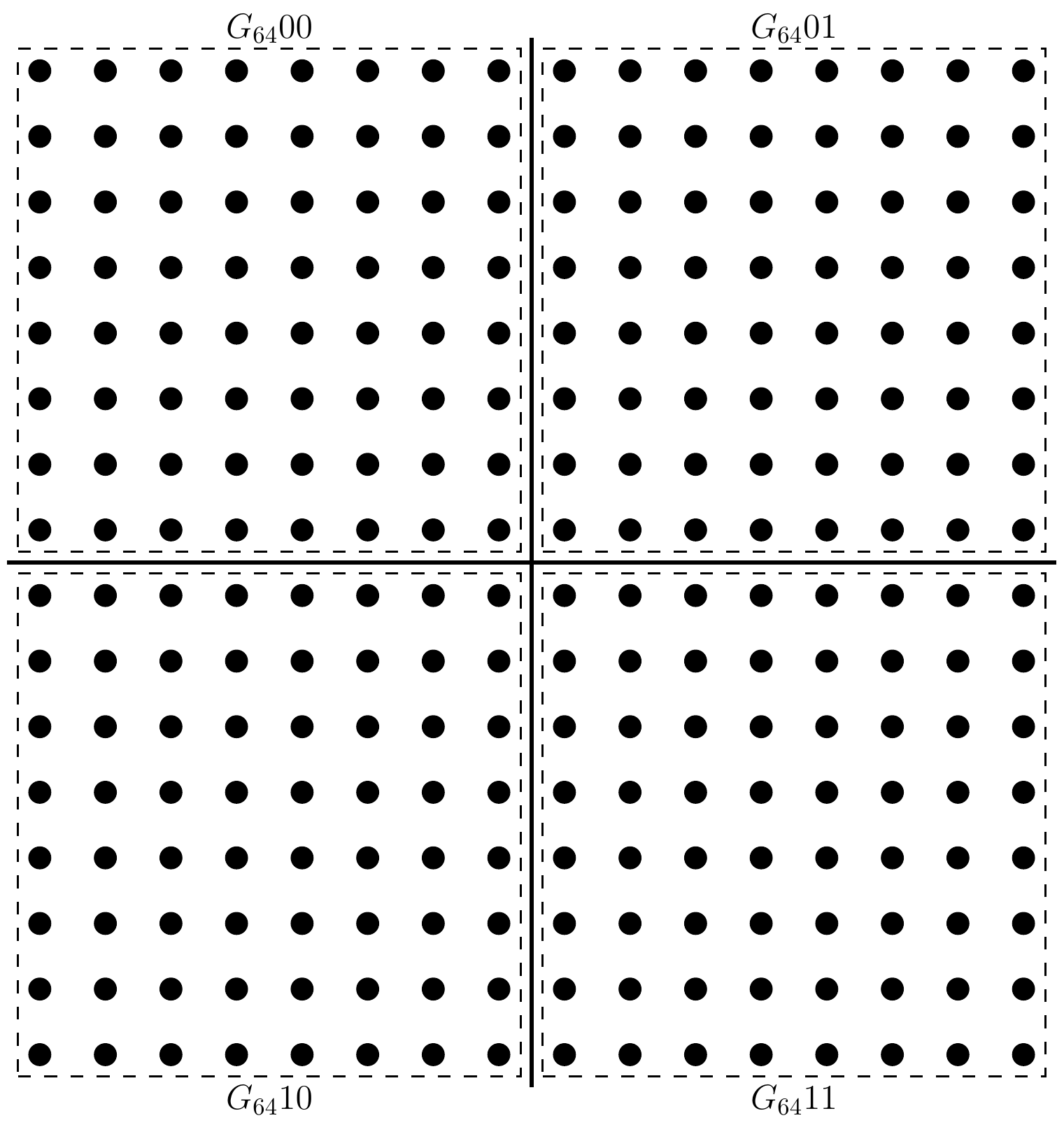}}
\fi
\caption{A mixed-labeled 256-QAM constellation, where $G_{64}$ represents a Gray-labeled 64-QAM constellation.} \label{fig:256}
\end{figure} 

In the following, we describe an adaptation of trellis
shaping~\cite{For1992a} to a bit-interleaved coded modulation system.
Consider the bit-labeling in Fig.~\ref{fig:256}.  For each point in a given
quadrant, the two most significant bits are the same; we refer to these two
bits as the \emph{shaping} bits. Furthermore, by the chain rule of mutual
information, we have
\begin{IEEEeqnarray}{rCl}
I(X;Y)&=&I(b_1,b_2,\ldots,b_{K},b_{K+1},b_{K+2};Y) \nonumber \\
&=&\underbrace{I(b_1,b_2,\ldots,b_{K};Y)}_{K\cdot R} \nonumber \\ 
& & \mbox{} +\underbrace{I(b_{K+1},b_{K+2};Y|b_1,b_2,\ldots,b_{K})}_{1},
\label{eq:shmlc} 
\end{IEEEeqnarray}
which provides a `two'-level (i.e., an MLC scheme with two generalized
levels) interpretation of the proposed system, with $M=K+2$. The first term
in (\ref{eq:shmlc}) represents the lowest level, to which error-correction
coding is applied; for the reasons stated previously, we will use
bit-interleaved coded modulation at this level. If the rate of the
error-correcting code is $R$, then this term communicates $K\cdot R$ bits
per symbol. The second term, the upper-level of the pseudo-MLC scheme, is
responsible for providing shaping. In trellis shaping, this is provided by
communicating---via $(b_{K+1},b_{K+2})$---a single bit per symbol, while
using a Viterbi-based shaping algorithm to select the remaining bit (of
freedom) to produce a sequence of symbols with a (nearly) bi-dimensional
Gaussian distribution.  Intuitively, the bi-dimensional Gaussian
distribution results from two facts: the Viterbi search selects the signal
path that minimizes energy, and for a fixed entropy, the Gaussian
distribution is the minimum energy distribution.

\begin{figure}[t]
\centering
\ifCLASSOPTIONonecolumn
\includegraphics[width=0.5\columnwidth]{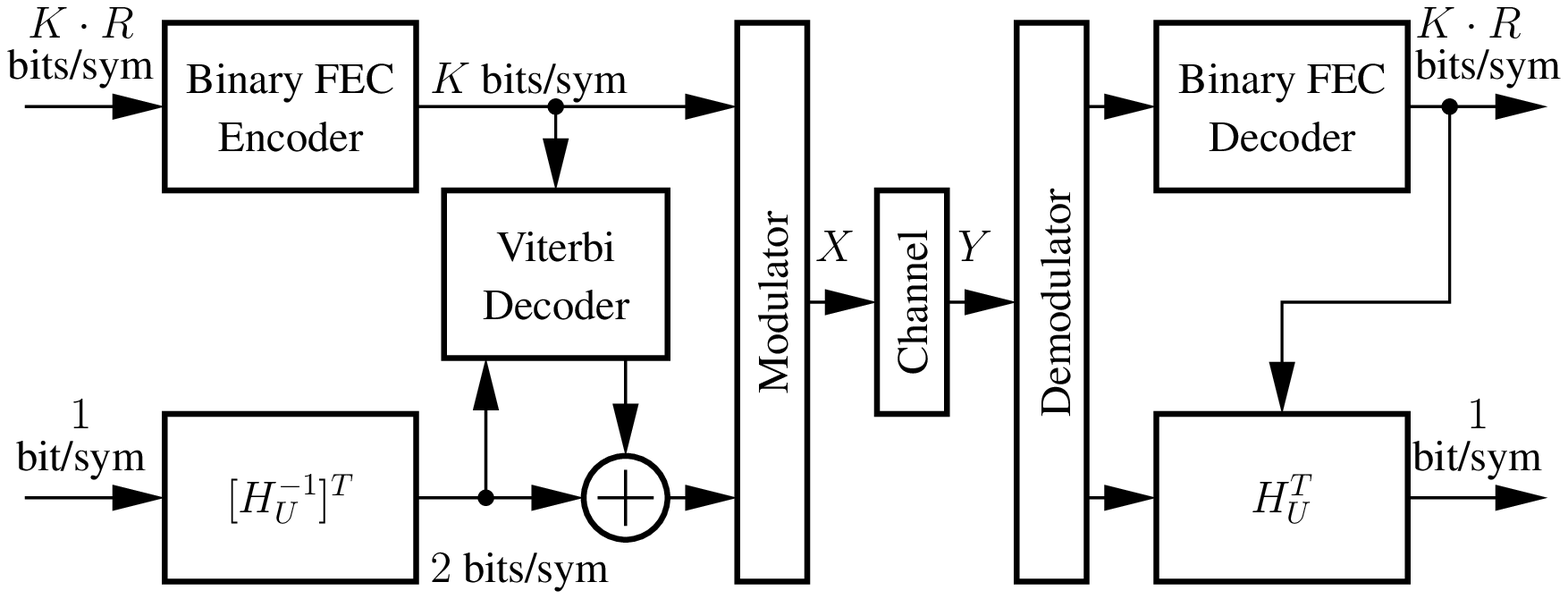}
\else
\includegraphics[width=\columnwidth]{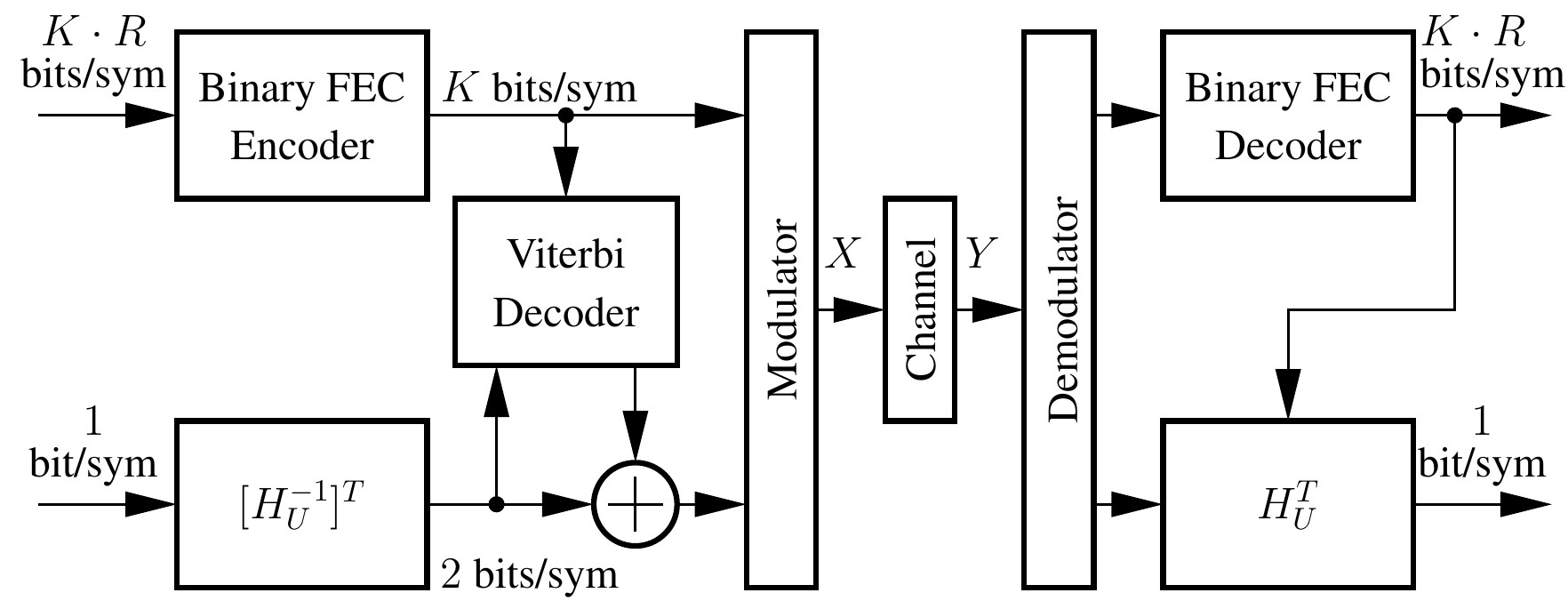}
\fi
\caption{The coded modulation system with shaping.} \label{fig:cdsys}
\end{figure}

As in~\cite{For1992a}, the Viterbi algorithm operates on the trellis of a
four-state convolutional code $C_U$ with generator matrix
$G_U=[1+D^2,1+D+D^2]$ and syndrome-former matrix $H_U^T=[1+D+D^2,1+D^2]^T$.
The overall operation of the system is as illustrated in
Fig.~\ref{fig:cdsys}.  

\subsection{Pragmatic Coded-Modulation via Staircase Codes}

To further reduce the complexity of the coded modulation system, we focus
on systems for which the error-correcting code (at the lowest level) is
decoded by a hard-decision decoder that receives hard decisions from the
channel.  That is, the demodulator in Fig.~\ref{fig:cdsys} outputs $K$ bits
(the $b_1,b_2,\ldots,b_{K}$ corresponding to the constellation point
closest to the received symbol) to the FEC decoder for every received
symbol; we assume coding is applied to these bits via BICM, and refer to
such a system as a ``pragmatic'' coded-modulation system.  

In a manner similar to that applied for the capacity estimates in
Fig.~\ref{fig:ramwdm}, the achievable rates of the pragmatic
coded-modulation system can be estimated via numerical simulations.  Since
BICM and hard-decision quantization are performed at the lowest level, the
capacity of the resulting level is 
\[
K(1-H_2(p_{\rm avg})),
\]
where $p_{\rm avg}$ is the average error rate of the received bits at the
lowest level, and $H_2(x) = -x \log_2 x - (1-x)\log_2(1-x)$ is the binary
entropy function.  Furthermore, the highest level communicates exactly 1
bit of information per symbol, and the maximum achievable information rate
for the pragmatic system is thus
\[
I_P=1+K(1-H_2(p_{\rm avg})).
\]
In Table~\ref{tab:rates}, the estimated values of $I_P$ are presented,
based on numerical simulations of the systems;  in each case, $K$ and the
average input power $P_{in}$ are optimized to maximize $I_P$. 

\begin{table}[t]
\centering
\caption{Achievable Rates Per Polarization
for Pragmatic Coded-Modulation System}
\begin{tabular}{lcccc} 
             &     &               & $P_{\rm in}$       & $I_P$ \\
Fiber System & $K$ & $p_{\rm avg}$ & (dBm) & (bits/s/Hz) \\ \hline
$L=500$~km, EQ & $8$ & $1.61\times 10^{-2}$ & $-6$ &  $8.05$\\
$L=500$~km, BP & $8$ & $3.52\times 10^{-3}$ & $-4$ &  $8.73$\\
$L=1000$~km, EQ & $6$ & $3.88\times 10^{-3}$ & $-6$ & $6.78$\\
$L=1000$~km, BP & $8$ & $2.22\times 10^{-2}$ & $-4$ & $7.77$\\
$L=2000$~km, EQ & $6$ & $2.52\times 10^{-2}$ & $-6$ & $5.98$\\
$L=2000$~km, BP & $6$ & $5.16\times 10^{-3}$ & $-4$ & $6.72$\\\hline
\end{tabular}
\label{tab:rates}
\end{table}

Note that for $L=2000$~km, the pragmatic coded modulations system has a
capacity within $0.47$ bits/s/Hz per polarization of the peak spectral
efficiency when only linear equalization is performed, and within $0.48$
bits/s/Hz per polarization for digital backpropagation.  For $L=1000$~km,
the pragmatic coded modulations system has a capacity within $0.62$
bits/s/Hz per polarization of the peak spectral efficiency when only linear
equalization is performed, and within $0.43$ bits/s/Hz per polarization for
digital backpropagation. Finally, for $L=500$~km, the pragmatic coded
modulations system has a capacity within $0.40$ bits/s/Hz per polarization
of the peak spectral efficiency when only linear equalization is performed,
and within $0.27$ bits/s/Hz per polarization for digital backpropagation.
In each case, the dominant contribution to the gap in performance is a
result of the hard quantization applied at the lowest level of the
(two-level) coded system.  Even though the hard quantization scheme leads
to some loss in performance, it is directly compatible with the
syndrome-based decoding of staircase codes. 

We now consider the design of staircase codes for use in the pragmatic
coded system.  In~\cite{Smi2011}, a G.709-compliant staircase code was
presented, with $R=239/255$, suitable for providing error-correction on a
binary symmetric channel with $p\leq 4.8 \times 10^{-3}$.  This code is
thus suitable for providing error-correction for the linearly-equalized
system with $L=1000$~km and the digitally-backpropagated system with
$L=500$~km.  For the other systems, we designed new staircase codes, the
parameters of which---including the net coding gain (NCG)---are provided in
Table~\ref{tab:scbicm}; the terminology used to describe the codes follows
that of Section~\ref{sec:sc}.

In each case, the length of the (mother) BCH component code is the smallest
$2^n-1$ that is greater than or equal to $2m$.

\begin{figure}[t] 
\centering
\ifCLASSOPTIONonecolumn
\includegraphics[width=0.5\columnwidth]{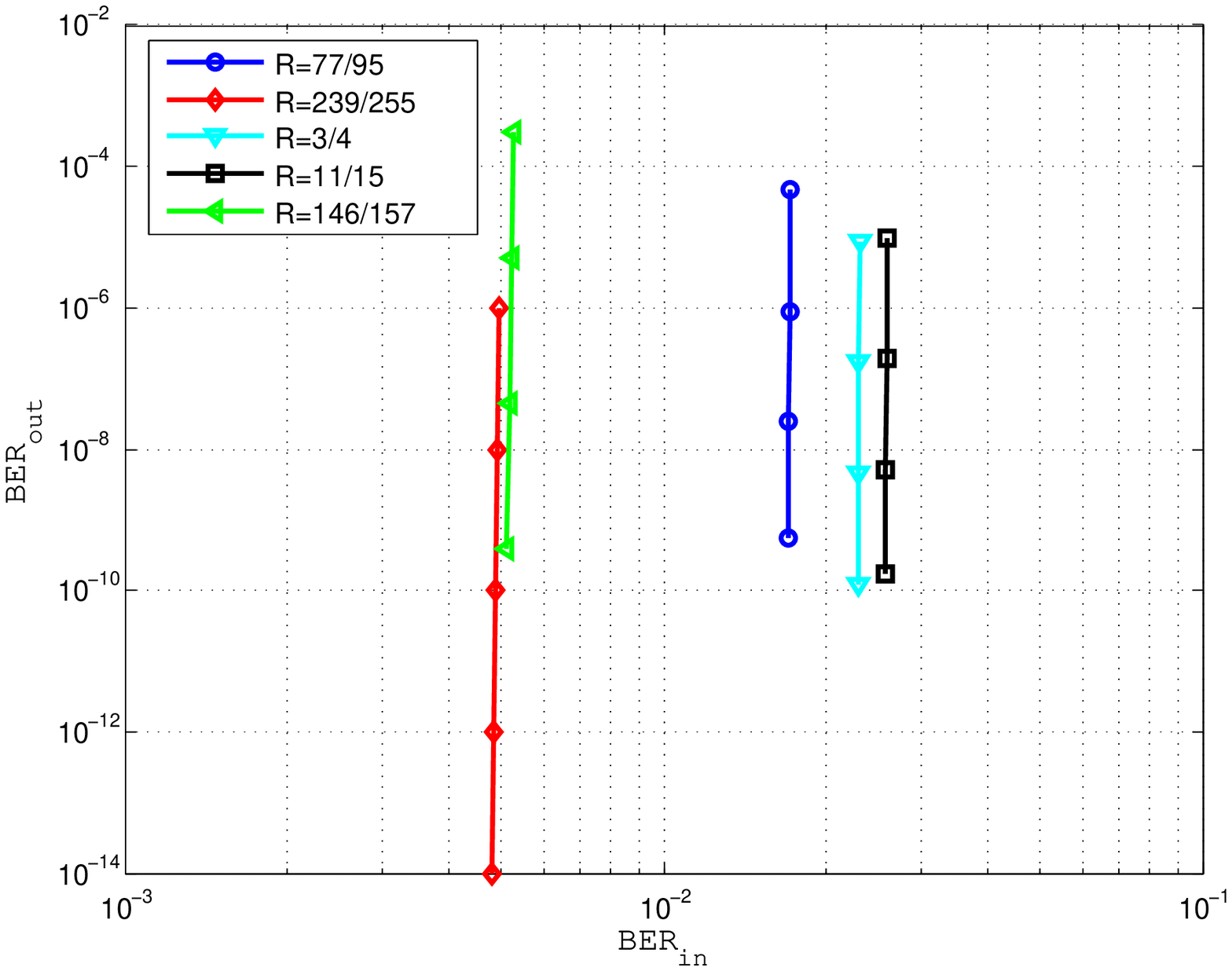}
\else
\includegraphics[width=0.950\columnwidth]{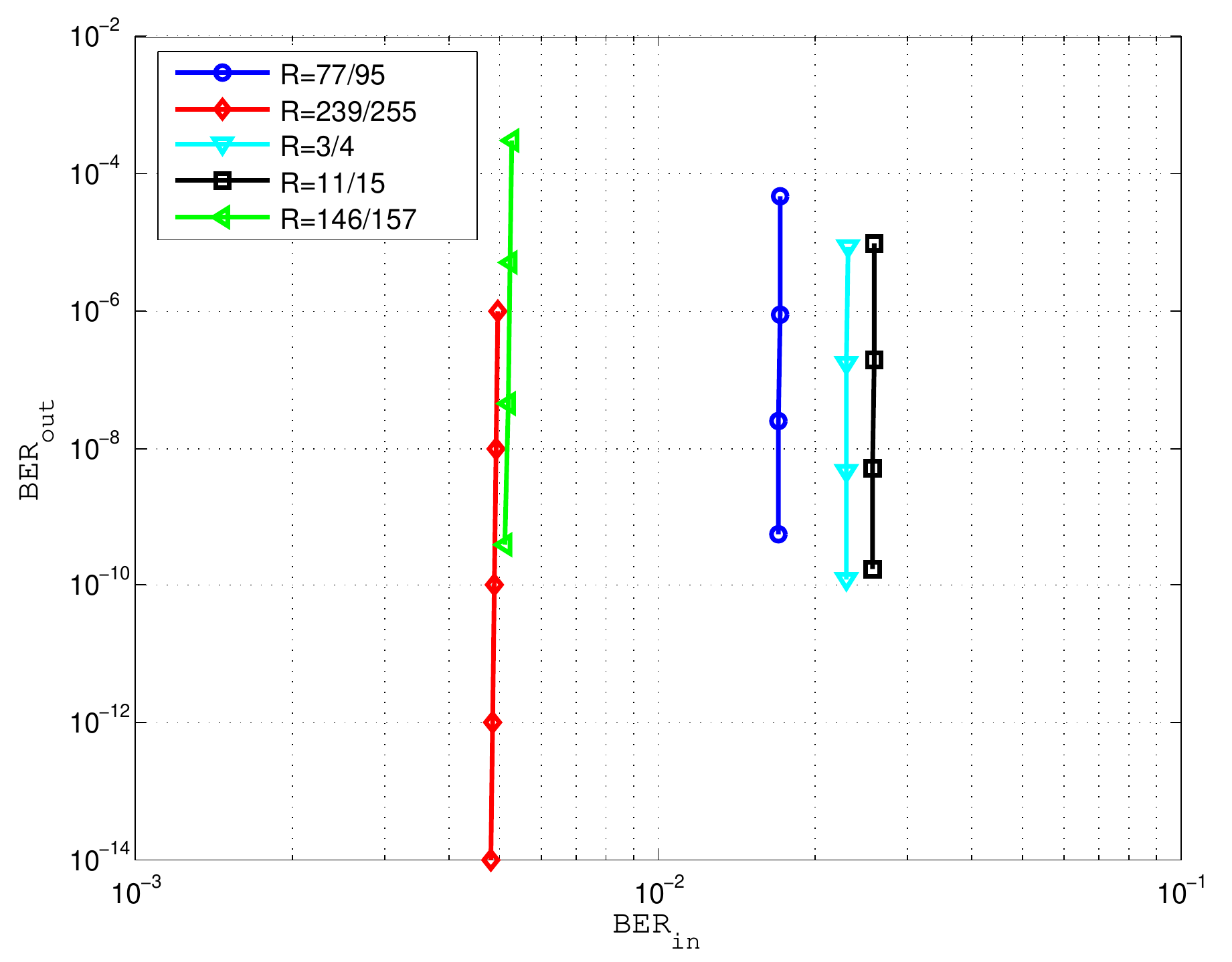}
\fi
\caption{Performance curves for the staircase codes in Table~\ref{tab:scbicm}.}
\label{fig:bersc} 
\end{figure}

In Fig.~\ref{fig:bersc}, the bit-error-rate curves are plotted.  Since
these curves (other than the G.709-compliant staircase code) were computed
without a hardware implementation, we were only able to obtain results to
approximately $10^{-10}$.  By the error floor estimation methods outlined
in~\cite{Smi2011} (with $p$ set to the average of the sub-channel error
rates), none of the systems have error floors above $10^{-20}$.  Thus,
extrapolating the curves to $10^{-15}$, each code has been designed to
provide an output error rate of better than $10^{-15}$ at the input error
rate induced by its corresponding system.  

In Fig.~\ref{fig:ramwdm}, the performance of the staircase coded systems is
plotted (the filled symbols), in addition to the estimated capacity curves
(the unfilled symbols). For $L=2000$~km, the system performs within $1.05$
bits/s/Hz per polarization of the peak spectral efficiency when only linear
equalization is performed, and within $0.62$ bits/s/Hz per polarization for
digital backpropagation.  For $L=1000$~km, the system performs within
$0.78$ bits/s/Hz per polarization of the peak spectral efficiency when only
linear equalization is performed, and within $1.2$ bits/s/Hz per
polarization for digital backpropagation. Finally, for $L=500$~km, the
system performs within $0.97$ bits/s/Hz per polarization of the peak
spectral efficiency when only linear equalization is performed, and within
$0.50$ bits/s/Hz per polarization for digital backpropagation.  

Note that the performance gap increases as the rate of the staircase code
(and corresponding sub-channel capacity) decreases, since staircase codes
perform closest to capacity at high rates.  The performance of those
systems could be improved by choosing a multi-dimensional constellation
that induces a higher rate (average) sub-channel. 

\begin{table}[ht]
\centering
\caption{Staircase Codes for Pragmatic Coded Systems}
\begin{tabular}{lccccc} 
             &     &     &     & NCG & Spec. Eff. \\ 
Fiber System & $m$ & $t$ & $R$ & (dB) & (bits/s/Hz) \\ \hline
$L=500$~km, EQ & $190$ & $4$ & $77/95$ &  10.47  & $7.48$\\
$L=500$~km, BP & $255$ & $3$ & $239/255$ &  9.41  & $8.50$\\
$L=1000$~km, EQ & $255$ & $3$ & $239/255$  &  9.41 & $6.62$\\
$L=1000$~km, BP & $144$ & $4$ & $3/4$ & 10.68  & $7.00$\\
$L=2000$~km, EQ & $120$ & $4$ & $11/15$ & 10.62  & $5.40$\\
$L=2000$~km, BP & $628$ & $4$ & $146/157$ &  9.50 & $6.58$\\\hline
\end{tabular}
\label{tab:scbicm}
\end{table}

\section{Conclusions}

We showed that digital backpropagation increases the achievable spectral
efficiencies---relative to linear equalization---by $0.55$ to $0.75$
bits/s/Hz per polarization.  We proposed a pragmatic coded modulation
system that incorporates signal shaping and exploits the excellent
performance and efficient high-speed decoding architecture of staircase
codes.  Reliable communication within $0.62$ bits/s/Hz per polarization of
the estimated capacity of a system with $L=2000$~km is provided by the
proposed system, with an error floor below $10^{-20}$.

\end{document}